\documentclass{article}

\usepackage{arXiv}

\usepackage[utf8]{inputenc} 
\usepackage[T1]{fontenc}    

\usepackage{xcolor}
\usepackage[colorlinks = true,
            linkcolor = gray,
            urlcolor  = gray,
            citecolor = gray,
            anchorcolor = red]{hyperref}
            
\usepackage{url}            

\usepackage{booktabs}       
\usepackage{multirow}       
\usepackage{tabularx}       
\usepackage{makecell}       
\newcolumntype{C}{>{\centering\arraybackslash}X} 

\usepackage{amsmath}        
\usepackage{amsfonts}       
\usepackage{nicefrac}       

\usepackage{microtype}      

\usepackage{apacite} 
\usepackage{doi}            

\usepackage{graphicx}       
\usepackage{float}          

\usepackage{fancyhdr}       
\title{Kilometer-Scale AI Downscaling of Atlantic Hurricanes with Generative Ensembles
    \author{
      \textbf{Yingkai Sha}$^{\dagger}$, %
      \textbf{Talea L. Mayo}$^{\ddagger}$,
      \textbf{Ethan D. Gutmann}$^{\dagger}$,
      \textbf{Lulin Xue}$^{\dagger}$,
      \textbf{Andrew Newman}$^{\dagger}$\\[1em]
      Research Applications Laboratory\\
      NSF National Center for Atmospheric Research, Boulder, Colorado, USA$^{\dagger}$\\[1em]
      Department of  Mathematics, Emory University, Atlanta, Georgia, USA$^{\ddagger}$
    }
}

\begin{document}
\maketitle

\begin{abstract}
This study presents an AI-based dynamical downscaling system for Tropical Cyclones (TCs). The system incorporates an AI-based limited-area model that downscales 3-hourly low-resolution boundary forcings into hourly high-resolution fields autoregressively, and a diffusion model that converts the outputs into ensembles of hazard-relevant variables. The system is trained on the regridded CONUS404 data with ERA5 forcings, and is evaluated on 20 TCs in 2020--2024. Verification shows stable downscaling performance across Atlantic hurricane seasons, with energy spectra closely matching the CONUS404 reference. The system is also verified to produce skillful TC-relevant weather extremes, largely improved over a deterministic AI baseline. The system performs well with forcing data from other models (GDAS/FNL) and can produce detailed eyewall, rainband, and landfall structures in TC case studies. The study provides a good example of how AI-based dynamical downscaling systems can be designed to resolve small-scale extreme weather events.
\end{abstract}

\vspace{2ex}
\textbf{Plain Language Summary}
Hurricanes cause damaging winds, heavy rainfall, and flooding, but global weather datasets are too coarse to describe these hazards in detail. To tackle this challenge, we developed a two-stage AI system that first reads coarse data and produces high-resolution weather over the southeastern United States, then connects it to a separate AI component that generates multiple possible versions of the high-resolution weather to characterize uncertainty. The two-stage system is tested on 20 Atlantic hurricanes in 2020--2024. The system ran stably, reproduced detailed hurricane structures, rainfall, and winds; it also worked well across multiple coarse datasets. This two-stage AI system is useful for supporting fast and high-quality hurricane weather hazard assessment.
\section{Introduction}

Tropical Cyclones (TCs) are a critical weather hazard, producing destructive winds, extreme rainfall, flooding, and storm surge \cite<e.g.>{emanuel2005increasing,knutson2020tropical}. Representing TCs requires high-resolution (HR) weather information beyond what low-resolution (LR) global models provide. Downscaling, which reconstructs HR weather from LR data, is therefore central to TC hazard assessment \cite{prein2015review,camargo2016tropical,lucas2021convection}. Dynamical downscaling runs storm-resolving regional models with LR boundary conditions to simulate HR weather \cite{prein2015review,rasmussen2023conus404}, effectively reproducing TC circulation and near-surface weather \cite{gutmann2018changes,roberts2020impact}. Statistical downscaling learns empirical LR-to-HR mappings from historical data \cite{maraun2010precipitation,gutmann2014intercomparison} and has long supported TC risk assessment \cite{emanuel2006statistical}. Snapshot-based AI downscaling implements such mappings with neural networks \cite{vandaldeepsd2017,sha2020adeep,sha2020bdeep,bano2020configuration}, including generative models \cite{harris2022generative,miralles2022downscaling}, to produce deterministic estimates and ensembles of HR fields, respectively, with successes in TC precipitation downscaling \cite{vosper2023deep} and stochastic TC field reconstruction \cite{mardani2025residual}.

Recent years have seen rapid progress in AI Weather Prediction \cite<AIWP; >{bi2023accurate,lam2023learning,chen2023fuxi,bodnar2024aurora,lang2024aifs,price2025probabilistic,schreck2025community}, and AI-based Limited-Area Models (LAMs) have extended this paradigm to finer regional grids \cite{nipen2025regional,xu2025artificial,abdi2025hrrrcast,pathak2026kilometer}. One emerging application is AI-based dynamical downscaling, in which the AI-based LAMs run continuously on seasonal and longer timescales under LR boundary forcings \cite{lopez2025dynamical,sha2026regional}. This approach evolves the full multivariate state within the HR space, preserving cross-variable relationships while enhancing spatiotemporal resolutions. It also offers efficient AI inference and couples naturally with generative models.

AI-based dynamical downscaling is promising for TCs because the continuously evolving HR state allows TCs to develop while the LR forcings supply only the large-scale environment. Few studies, however, have examined this application. To fill this gap, this study develops and evaluates a two-stage AI-based dynamical downscaling system for TCs and their associated weather extremes. In detail, an AI-based LAM is proposed first to downscale 3-hourly LR boundary forcings into hourly HR fields over a southeastern CONUS domain, then a generative model converts these deterministic outputs into ensembles of hazard-relevant variables. The system is trained on the convection-permitting, 40-year, 4-km, CONUS domain simulation \cite<CONUS404; >{rasmussen2023conus404}, aggregated to 8 km grids, with the ECMWF Reanalysis version 5 \cite<ERA5; >{hersbach2020era5} serving as the LR forcing. The evaluation covers 20 named Atlantic TCs in 2020--2024 and includes non-ERA5 forcings to test generalization.

This study addresses three research questions: (1) How does the system perform overall during the Atlantic hurricane season? (2) To what extent can it reconstruct the fine-scale variability and weather extremes? (3) Does it generalize to boundary forcings beyond its training datasets? To answer these questions, we aim to establish an AI-based workflow for TC downscaling at storm-resolving scales, quantify its effectiveness for extreme weather events, and inform future AIWP development for TC risk assessment.

\section{Research domain and data}\label{sec2}
\subsection{Region of interest}\label{sec21}

\begin{figure}
    \centering
    \includegraphics[width=0.9\columnwidth]{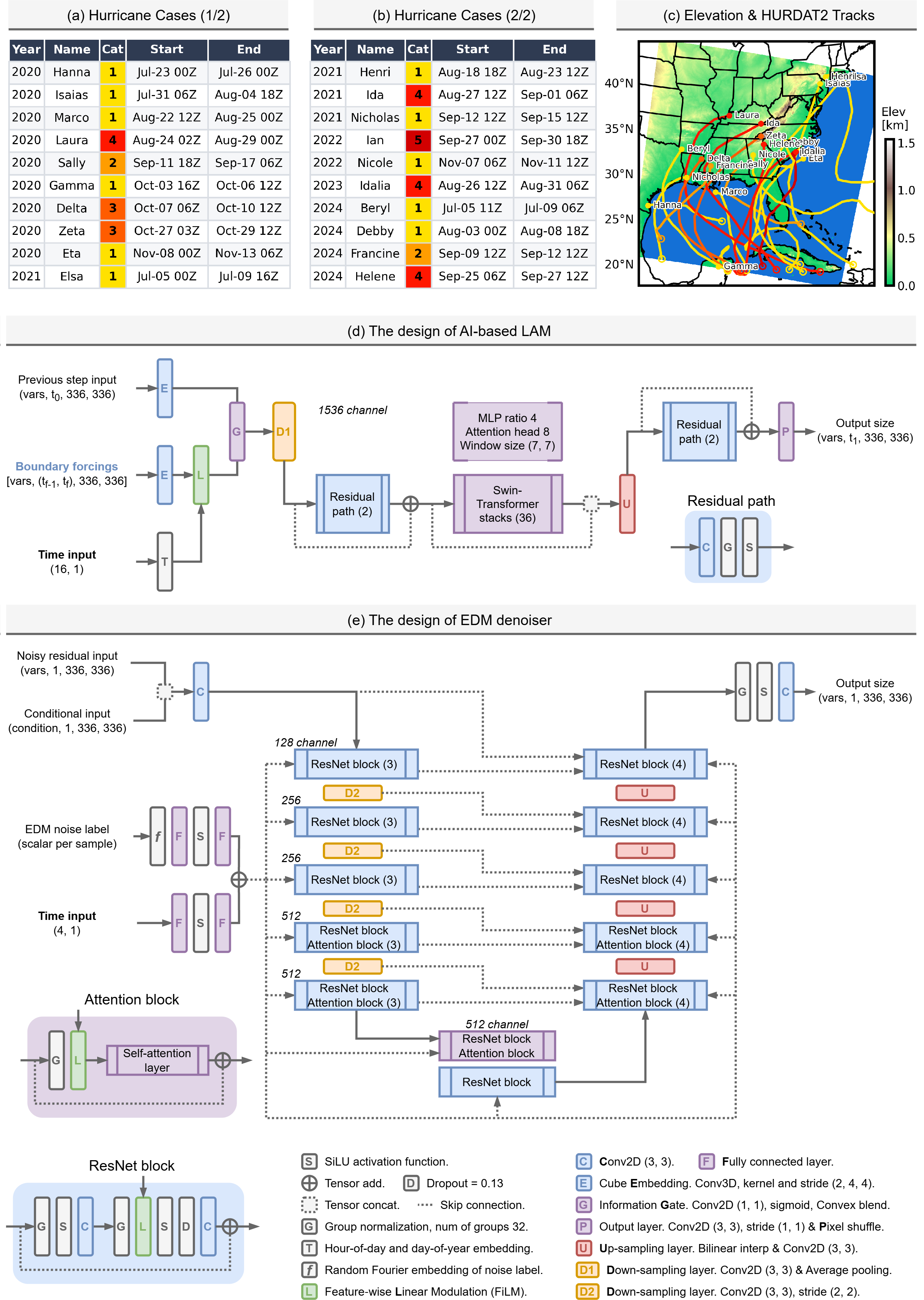}
    \caption{
    (a) and (b) The name, category, and duration of 20 TCs evaluated in this study. 
    (c) The coverage of the 8-km domain with shaded elevation and TC tracks.
    (d) Architecture of the AI-based LAM.
    (e) Architecture of the EDM denoiser, serving as the diffusion model.
    }
    \label{fig1}
\end{figure}

This study focuses on AI-based dynamical downscaling over the Gulf and the southeastern United States, spanning 18--45$^\circ$N and 65--99$^\circ$W with 8-km grid spacing (Figure~\ref{fig1}c). This domain covers the primary landfall regions of Atlantic TCs. During the hurricane season, the warm waters of the Gulf benefit TC development and intensification; the guiding flow on the western edge of the North Atlantic subtropical high directs many TCs toward the Gulf and Southeast coast, making this region the destination of many TC landfalls \cite<e.g.>{klotzbach2018continental}. The domain also contains the southern Appalachians (Figure~\ref{fig1}c), whose terrain modulates inland TC rainfall and decay. Within the 2020--2024 hurricane seasons, 20 named TCs are identified in this domain and thus selected for evaluation (Figure~\ref{fig1}a,b); they span Saffir--Simpson Categories 1--5, including Hurricane Ian, which caused more than 150 fatalities and over \$112 billion in damage \cite{bucci2023ian}. Thus, the dynamical downscaling over this domain is scientifically meaningful for demonstrating the ability of the AI system to reproduce TC hazards, and is of high priority for disaster risk management.

\subsection{Training Data}\label{sec22}
\begin{table}
\centering
\caption{Summary of inputs and outputs in this study.}\label{tab1}
\renewcommand{\arraystretch}{1.2}
\begin{tabularx}{\textwidth}{c c >{\centering\arraybackslash}X c c}
\specialrule{1.5pt}{0pt}{3pt}
Usage & Type & Variable name & Units & Role \\
\midrule
\multirow{12}{*}{\makecell{AI-based LAM \\ 8-km input \\ and output}}
& \multirow{5}{*}{Upper air\textsuperscript{a}}
  & Zonal wind               & $\mathrm{m \cdot s^{-1}}$   & \multirow{5}{*}{\makecell{Input and \\ output}} \\
& & Meridional wind          & $\mathrm{m \cdot s^{-1}}$   & \\
& & Air temperature          & $\mathrm{K}$                & \\
& & Specific humidity        & $\mathrm{kg \cdot kg^{-1}}$ & \\
& & Total pressure           & $\mathrm{Pa}$               & \\
\cmidrule(lr){2-5}
& \multirow{5}{*}{Single level}
  & Surface pressure         & $\mathrm{Pa}$              & \multirow{5}{*}{\makecell{Input and \\ output}} \\
& & 2-Meter temperature      & $\mathrm{K}$               & \\
& & 10-Meter zonal wind      & $\mathrm{m \cdot s^{-1}}$  & \\
& & 10-Meter meridional wind & $\mathrm{m \cdot s^{-1}}$  & \\
& & Total precipitable water & $\mathrm{kg \cdot m^{-2}}$ & \\
\cmidrule(lr){2-5}
& \multirow{2}{*}{Static\textsuperscript{c}}
  & Elevation                & $\mathrm{m}$ & \multirow{2}{*}{Input only} \\
& & Land--sea mask           & --           & \\
\midrule
\multirow{9}{*}{\makecell{AI-based LAM \\ boundary \\ forcing inputs}}
& \multirow{4}{*}{Upper air\textsuperscript{b}}
  & Zonal wind        & $\mathrm{m \cdot s^{-1}}$   & \multirow{4}{*}{Input only} \\
& & Meridional wind   & $\mathrm{m \cdot s^{-1}}$   & \\
& & Air temperature   & $\mathrm{K}$                & \\
& & Specific humidity & $\mathrm{kg \cdot kg^{-1}}$ & \\
\cmidrule(lr){2-5}
& \multirow{5}{*}{Single level}
  & Mean sea level pressure  & $\mathrm{Pa}$              & \multirow{5}{*}{Input only} \\
& & 2-Meter temperature      & $\mathrm{K}$               & \\
& & 10-Meter zonal wind      & $\mathrm{m \cdot s^{-1}}$  & \\
& & 10-Meter meridional wind & $\mathrm{m \cdot s^{-1}}$  & \\
& & Total precipitable water & $\mathrm{kg \cdot m^{-2}}$ & \\
\midrule
\multirow{6}{*}{\makecell{Diffusion model \\ input and output \\ variables\textsuperscript{c}}}
& \multirow{6}{*}{\makecell{Single level}}
  & Surface pressure         & $\mathrm{Pa}$              & \multirow{5}{*}{\makecell{Input and \\ output}} \\
& & 2-Meter temperature      & $\mathrm{K}$               & \\
& & 10-Meter zonal wind      & $\mathrm{m \cdot s^{-1}}$  & \\
& & 10-Meter meridional wind & $\mathrm{m \cdot s^{-1}}$  & \\
& & Total precipitable water & $\mathrm{kg \cdot m^{-2}}$ & \\
\cmidrule(lr){3-5}
& & Total precipitation & $\mathrm{mm}$ & Output only \\
\specialrule{1.5pt}{3pt}{0pt}
\end{tabularx}

\vspace{2pt}
\raggedright
\textsuperscript{a}\,Upper-air variables of the AI-based LAM are on 12 hybrid sigma--pressure levels.\\
\textsuperscript{b}\,Upper-air variables of the LR boundary forcings are on 6 constant pressure levels.\\
\textsuperscript{c}\,Static variables are also used as diffusion model inputs.\\
\end{table}

The AI-based LAM is trained on paired HR targets and LR boundary forcings over the region of interest. The HR data are obtained from the CONUS404 \cite{rasmussen2023conus404}, aggregated from its native 4-km grid to the 8-km domain. The \{0, 3, 6, 9, 12, 15, 18, 21, 24, 30, 36, 42\}-th terrain-following Weather Research and Forecasting (WRF) model levels are selected. The 8-km re-gridding would not lose key information because the intrinsic resolution of the numerics in WRF is lower than its native 4-km grid \cite{skamarock2004evaluating}. The LR boundary forcings are from ERA5 \cite{hersbach2020era5} on the \{950, 900, 850, 700, 500, 200\}~hPa constant pressure levels, together with the single-level fields (Table~\ref{tab1}), and are interpolated linearly from 0.25$^\circ$ to the 8-km grids.

A generative model is proposed as the second stage of the AI-based dynamical downscaling system. This generative stage reconstructs fine-scale HR details that deterministic training tends to smooth, while producing ensembles of hazard-relevant variables. A diffusion model inspired by \citeA{karras2022elucidating} and \citeA{mardani2025residual} is selected to produce ensembles of key single-level variables (Table~\ref{tab1}). Variables except total precipitation are refined as paired inputs and outputs, whereas total precipitation, which the previous stage AI-based LAM does not predict, is generated as an output-only variable.

The pre-processed CONUS404 and ERA5 are subset to 1980--2018 for training and the year 2019 for validation. ERA5 is downsampled from hourly to 3-hourly for the training of the AI-based LAM. Following \citeA{sha2026regional}, 3-hourly boundary steps are selected as a practical optimum that balances generalization ability and downscaling accuracy. Variables are normalized using z-scores followed by residual normalization \cite{watt2023ace,sha2025improving,sha2025investigating}. Specific humidity and total precipitable water are square-root transformed before z-scoring, and total precipitation is quad-root transformed. All coefficients are computed from the training subset. The land--sea mask is converted to binary values of 0 and 1.

\subsection{Testing and Verification Data}\label{sec23}

The AI-based dynamical downscaling system is tested with multiple boundary forcing datasets covering the 2020--2024 hurricane seasons (1 June to 30 November of each year), except for the year 2024, for which the CONUS404 data ends on 30 September 2024. The ERA5 at 3-hourly steps is applied, serving as the LR forcing for evaluating the hourly downscaling performance of the system. The Global Data Assimilation System/Final Analysis \cite<GDAS/FNL; >{NCEP2015FNL} is also selected as an unseen test forcing independent of ERA5, as it is produced by the data assimilation component of the Global Forecast System (GFS). Skillful downscaling of GDAS/FNL forcing would not only highlight the generalization ability of the AI-based dynamical downscaling system but also indicate the future potential of applying the system to other GFS products. Here, GDAS/FNL is provided at 3-hourly, 0.25$^\circ$ resolution, and is interpolated to the 8-km grid in the same way as ERA5. For verification, the regridded CONUS404 serves as the HR reference; the TC positions and intensities from the National Hurricane Center Hurricane Database \cite<HURDAT2; >{landsea2013atlantic} are applied to evaluate the 20 selected TCs.

\section{Methods}\label{sec3}
\subsection{Model design and training}\label{sec31}

The AI-based LAM follows the Swin-Transformer-based design of \citeA{sha2026regional} (Figure~\ref{fig1}d). It encodes the previous-step HR state and the boundary forcings at two neighboring times through separate branches, injects the encoded; time information through Feature-wise Linear Modulation \cite<FiLM; >{perez2018film}, and blends the two streams with a gated fusion before the Swin-Transformer core. The output layer maps the features back to the HR grid through pixel shuffle \cite{shi2016real}, producing the next-hour state that is recycled autoregressively.

The generative stage employs a diffusion model based on the ``Elucidating the Design Space'' \cite<EDM; >{karras2022elucidating} formulation, as adapted in CorrDiff \cite{mardani2025residual} (Figure~\ref{fig1}e). The EDM denoiser is conditioned on selected AI-based LAM outputs and static fields (Table~\ref{tab1}); it employs self-attention \cite{ho2020denoising}, and is modulated through FiLM by embeddings of the noise configuration and time inputs.

The AI-based LAM is trained from scratch, whereas the diffusion model is trained on the hindcast of the AI-based LAM over the training period of 1980--2018. Both models use the AdamW optimizer \cite{loshchilov2017decoupled,wang2024set} with a weight decay of $3\times 10^{-6}$, 32 samples per batch, and initial learning rates of $10^{-3}$ and $10^{-4}$, respectively, with half-cosine annealing. The AI-based LAM training follows \citeA{sha2026regional}. Its single-step pre-training runs 120 epochs with 3000 batches per epoch, and multi-step fine-tuning then applies a fixed learning rate of $3\times 10^{-7}$ over 5 epochs, extending the rollout to 2--6 steps. The diffusion model takes the AI-based LAM outputs as conditional input and its CONUS404 residual as targets. Its training runs 90 epochs with 2000 batches per epoch, using the noise-level-weighted loss of \citeA{karras2022elucidating} with log-normally drawn noise levels.

All models are implemented in PyTorch \cite{paszke2019pytorch} using the Community Research Earth Digital Intelligence Twin \cite<CREDIT; >{schreck2025community} platform at the NSF National Center for Atmospheric Research (NCAR). The single-step and multi-step training of the AI-based LAM, and the diffusion training take approximately 180, 60, and 100 wall-clock hours, respectively, on 32 NVIDIA A100 GPUs. 
For inference, the AI-based LAM and the diffusion model take roughly 1 and 5 GPU-hours, respectively, on a single GPU, to produce a single member for one entire Atlantic hurricane season.
Technical details of the AI model training, inference, and hyperparameters are provided in the Supporting Information.
 
\subsection{Experiment design}\label{sec32}

This study runs three downscaling experiments:
\begin{enumerate}
    \item \textit{LAM-ERA5} covers the 2020--2024 Atlantic hurricane seasons and uses 3-hourly ERA5 as boundary forcings. This experiment evaluates the performance of the AI-based dynamical downscaling system and provides a ``no generalization'' reference for other experiments.
    
    \item \textit{LAM-GDAS} covers the same seasons and replaces ERA5 with 3-hourly, 0.25$^\circ$ GDAS/FNL. Its comparison with LAM-ERA5 examines how well the AI-based dynamical downscaling system generalizes to datasets that it was not trained on.
 

    \item \textit{Sha et al.~(2020) baseline} covers the same seasons and replicates the deterministic, snapshot-based AI downscaling of \citeA{sha2020adeep} on hourly ERA5. Its UNet design has been widely adopted in later downscaling studies \cite{wang2021deep,sun2024deep,hsieh2023deep} and thus represents conventional AI downscaling approaches well. This baseline serves as a reference for quantifying the added value of the two-stage system.
\end{enumerate}
\noindent
LAM-ERA5 and LAM-GDAS are initialized 3 days prior to the Atlantic hurricane season each year using 2010--2019 CONUS404 climatology in May, allowing the system to spin up. LAM-ERA5 and LAM-GDAS produce 20 generative members for evaluation.

\subsection{Verification methods}\label{sec33}

Downscaling skills across the Atlantic hurricane seasons are verified with the Mean Absolute Error (MAE) for deterministic outputs, and the Continuous Ranked Probability Score \cite<CRPS; >{hersbach2000decomposition,gneiting2007strictly} for the ensembles. CRPS reduces to MAE for deterministic predictions, so the two scores are directly comparable, and an ensemble CRPS below the deterministic MAE indicates added downscaling accuracy and probabilistic value. The restoration of fine-scale variability is evaluated using azimuthally averaged energy spectra \cite{skamarock2004evaluating,morss2009spectra}, compared against the CONUS404 reference to diagnose the effective resolution of the downscaling outputs.

TC-specific performance is verified against the HURDAT2 observations \cite{landsea2013atlantic}. The central intensity is evaluated as the maximum 10-m wind speed near the center, and the absolute track error is the great-circle distance between the downscaled and observed centers at matched times.

10-m wind and precipitation extremes are verified as threshold exceedances using the Symmetric Extremal Dependence Index \cite<SEDI; >{ferro2011extremal}, the Equitable Threat Score \cite<ETS; >{schaefer1990critical}, and the Fractions Skill Score \cite<FSS; >{roberts2008scale}.
The calibration of ensemble-derived exceedance probabilities is examined with reliability diagrams \cite{wilks2019statistical}. All extreme-value scores are computed only at times when at least one TC is present in the domain, focusing the evaluation on TC-relevant extremes. Technical details of the verifications are provided in Supporting Information.

\section{Results}
\subsection{Verification of overall performance}

\begin{figure}
    \centering
    \includegraphics[width=\columnwidth]{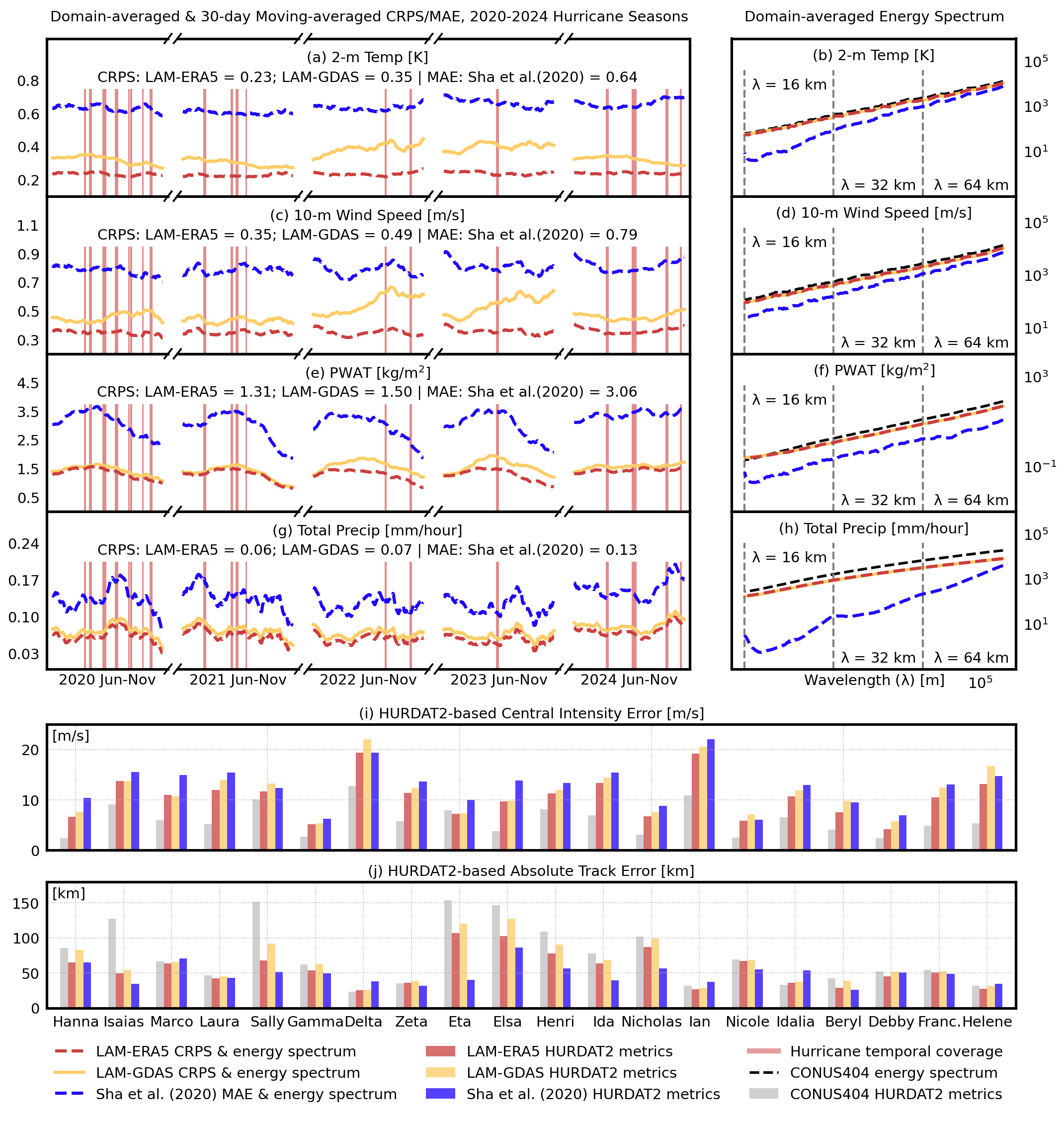}
    \caption{
    (a) Domain-averaged and 30-day moving-averaged hourly 2-m air temperature (``2-m Temp'') CRPS for LAM-ERA5 (red) and LAM-GDAS (orange), and MAE for the Sha et al.~(2020) baseline (blue) in the 2020--2024 Atlantic hurricane seasons. Light red shading indicates the temporal coverage of TCs. 
    (b) The energy spectrum of 2-m air temperature in the 10--100 km wavelength range, where the black dashed line is computed from CONUS404. 
    (c--d) As in (a--b), but for 10-m wind speed. 
    (e--f) As in (a--b), but for total precipitable water (``PWAT''). 
    (g--h) As in (a--b), but for total precipitation. 
    (i--j) The HURDAT2-based TC central intensity and absolute track errors, respectively, for CONUS404 (gray), LAM-ERA5 (red), LAM-GDAS (orange), and the Sha et al.~(2020) baseline (blue). TC names are listed from 2020 (left) to 2024 (right).
    Note that the year 2024 result ends on 30 September 2024.}
    \label{fig2}
\end{figure}

All downscaling experiments are verified over the 2020--2024 Atlantic hurricane seasons (Figure~\ref{fig2}). The LAM-ERA5 and LAM-GDAS maintain stable CRPS with no error spikes (Figure~\ref{fig2}a, c, e, and g), confirming that the system iterates stably over season-long integrations. LAM-GDAS is comparable to LAM-ERA5 for total precipitable water and total precipitation and moderately worse for 2-m air temperature and 10-m wind speed. Although trained with ERA5 only, the system generalizes well to the unseen GDAS/FNL forcings; this finding is consistent with \citeA{sha2026regional}. 

The LAM-ERA5 and LAM-GDAS produce CRPS values roughly half of the MAE of the Sha et al.~(2020) baseline. This margin indicates that the AI-based dynamical downscaling system adds clear value over conventional AI downscaling. A mild within-season increase of the LAM-GDAS 10-m wind CRPS appears in 2022 and 2023, but the scores remain well below the baseline errors. Pure MAE-based verifications, computed from 20-member averages and from the AI-based LAM outputs, are provided in the Supporting Information. They are consistent with the findings in Figure~\ref{fig2}.

The domain-averaged energy spectra of LAM-ERA5 and LAM-GDAS closely track the regridded CONUS404 reference, whereas the Sha et al.~(2020) baseline exhibits a variance deficit at fine scales, especially for total precipitation (Figure~\ref{fig2}h). The spectra also improve upon those reported for the AI-based LAM alone in \citeA{sha2026regional}, indicating that the generative stage contributes to the two-stage system by restoring spatial variance on storm-resolving scales.

CONUS404 and all predictions have TC intensity errors (Figure~\ref{fig2}i) because the gridded 10-m wind cannot fully represent the point-scale, one-minute averaged wind maxima of the HURDAT2. The errors typically grow with TC intensity and peak for strong cases such as Ian. CONUS404 exhibits the lowest errors (except Eta) and marks the representativeness floor, while LAM-ERA5 and LAM-GDAS add moderate errors above CONUS404, outperforming Sha et al.~(2020) baseline for 20 and 15 out of 20 TCs, respectively. This indicates that the AI-based dynamical downscaling system can preserve adequate central intensity information.

For the absolute track errors (Figure~\ref{fig2}j), the two ERA5-driven results, LAM-ERA5 and the Sha et al.~(2020) baseline, are slightly better overall, whereas CONUS404 exhibits the largest errors for several TCs (e.g., Sally, Eta, and Elsa). This behavior reflects how TC positions are influenced by large-scale environments. ERA5 and GDAS/FNL assimilate observations, so their TC positions, inherited by the downscaling experiments through the LR forcings, stay close to the HURDAT2 tracks. CONUS404, in contrast, is a dynamical downscaling simulation constrained by lateral boundaries and spectral nudging only; this allows its TCs to drift from the observed tracks. Combining the HURDAT2-based TC verifications, the AI-based dynamical downscaling system makes good use of the forcing-guided TC positions while also generating CONUS404-like fine-scale spectra; although trained on CONUS404, the system is not bound to the track deficits of its training target.

\subsection{Verifications of TC extremes}

\begin{figure}
    \centering
    \includegraphics[width=\columnwidth]{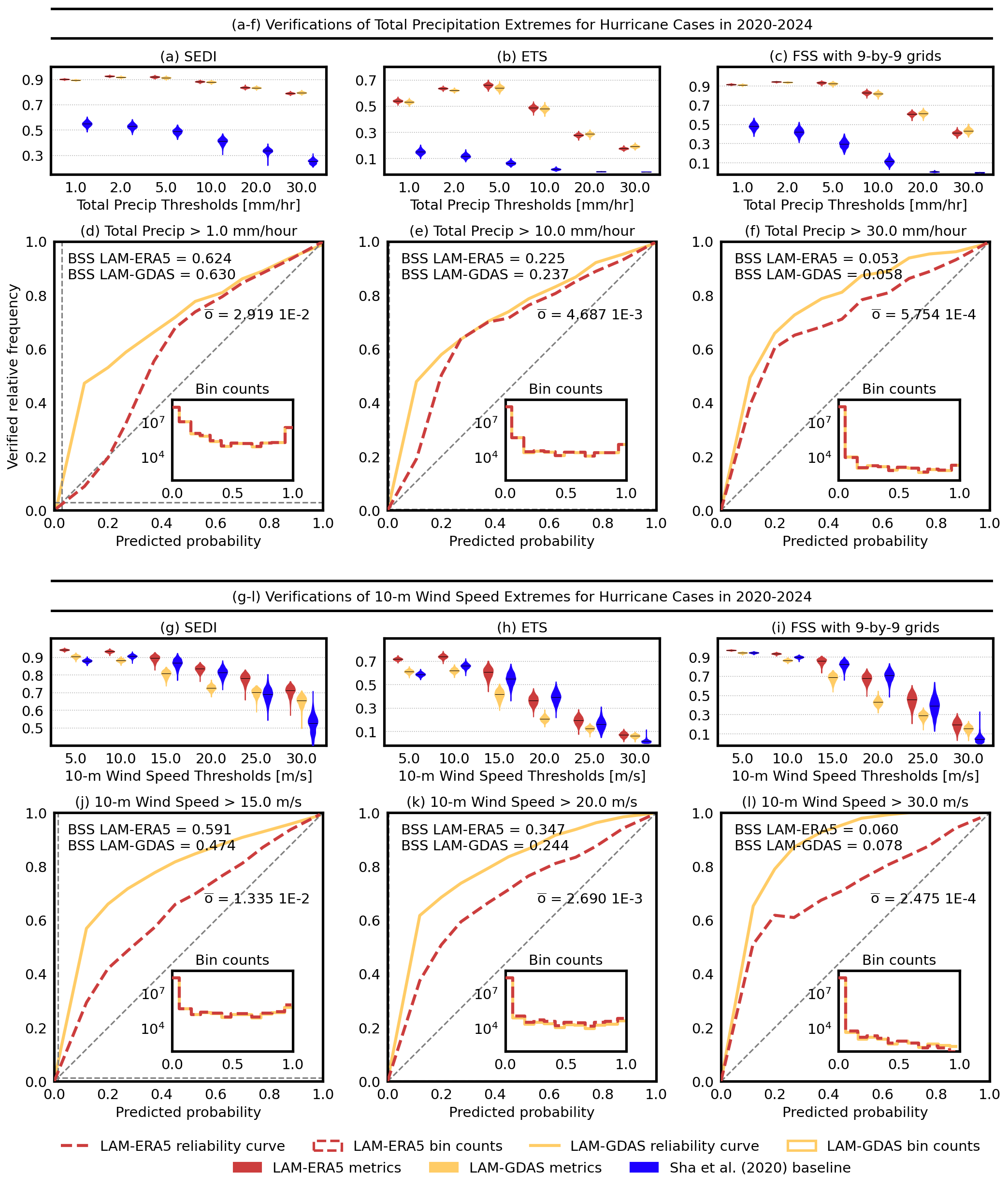}
    \caption{
    Threshold-based verification of total precipitation extremes across the temporal coverage of all 20 TCs in 2020--2024. 
    (a--c) Scores (y-axis) of SEDI, ETS, and FSS with 9-by-9 grids vary by total precipitation thresholds (x-axis), respectively for LAM-ERA5 (red), LAM-GDAS (orange), and Sha et al.~(2020) baseline (blue). Violin markers represent scores of individual ensemble members or deterministic outputs, with 1000-step bootstrapping. 
    (d--f) Reliability diagrams of total precipitation exceeding 1.0, 10.0, and 30.0 mm per hour rate, respectively, for LAM-ERA5 (red) and LAM-GDAS (orange). Each diagram has predicted probability as the x-axis, verified relative frequency as the y-axis, and bin counts in the subpanel. Brier Skill Scores (BSS) and the uncertainty term ($\overline{o}$) values are provided.
    (g--l) As in (a--f), but for 10-m wind speed extreme events.}
    \label{fig3}
\end{figure}

For total precipitation extremes, LAM-ERA5 and LAM-GDAS are verified to outperform the Sha et al.~(2020) baseline at all thresholds (Figure~\ref{fig3}a--c). Their SEDI are high and stable, whereas the Sha et al.~(2020) baseline decays rapidly; FSS shows a similar separation, with the baseline losing nearly all skill beyond 10 $\mathrm{mm\cdot hour^{-1}}$. The ETS of LAM-ERA5 and LAM-GDAS peaks at 5 $\mathrm{mm\cdot hour^{-1}}$, slightly decays at higher thresholds, but still outperforms the Sha et al.~(2020) baseline. The near-identical LAM-ERA5 and LAM-GDAS scores indicate that the system can produce skillful precipitation extremes, and this benefit generalizes well to the unseen GDAS/FNL forcing.

The 10-m wind verifications are more competitive (Figure~\ref{fig3}g--i). LAM-ERA5 and the Sha et al.~(2020) baseline are comparable at thresholds up to 20 m~s$^{-1}$, reflecting that near-surface wind extremes inherit more directly from the LR circulation than small-scale precipitation. LAM-ERA5 leads at the thresholds of 25 and 30 m~s$^{-1}$. LAM-GDAS falls behind both ERA5-based experiments between 15 and 30 m~s$^{-1}$; this can be explained by the quality difference of horizontal winds in ERA5 and GDAS/FNL \cite{li2024exploring}. The 15--30 m~s$^{-1}$ range also covers the wind speed thresholds that define TC wind radii, so skill in this range reflects how well each experiment captures the extent of TC-forced winds. This wind speed also affects the integrated kinetic energy, which highlights surge potential better than maximum intensity alone \cite{powell2007tropical}.

Reliability diagrams are verified in Figures~\ref{fig3}d--f and j--l. The Brier Skill Score (BSS) remains positive at all thresholds, indicating that the downscaled ensembles are overall usable for these extremes. The reliability curves, however, tend to sit above the identity line, suggesting that the generative stage is not fully calibrated. This is recognized as a common issue for generative AI approaches since it has been reported from the CorrDiff study \cite{mardani2025residual}. Given that the reliability curves cover both low- and high-probability bins with adequate samples, we think this miscalibration can be addressed by other downstream methods in the future. 

\subsection{Case studies}

\begin{figure}
    \centering
    \includegraphics[width=\columnwidth]{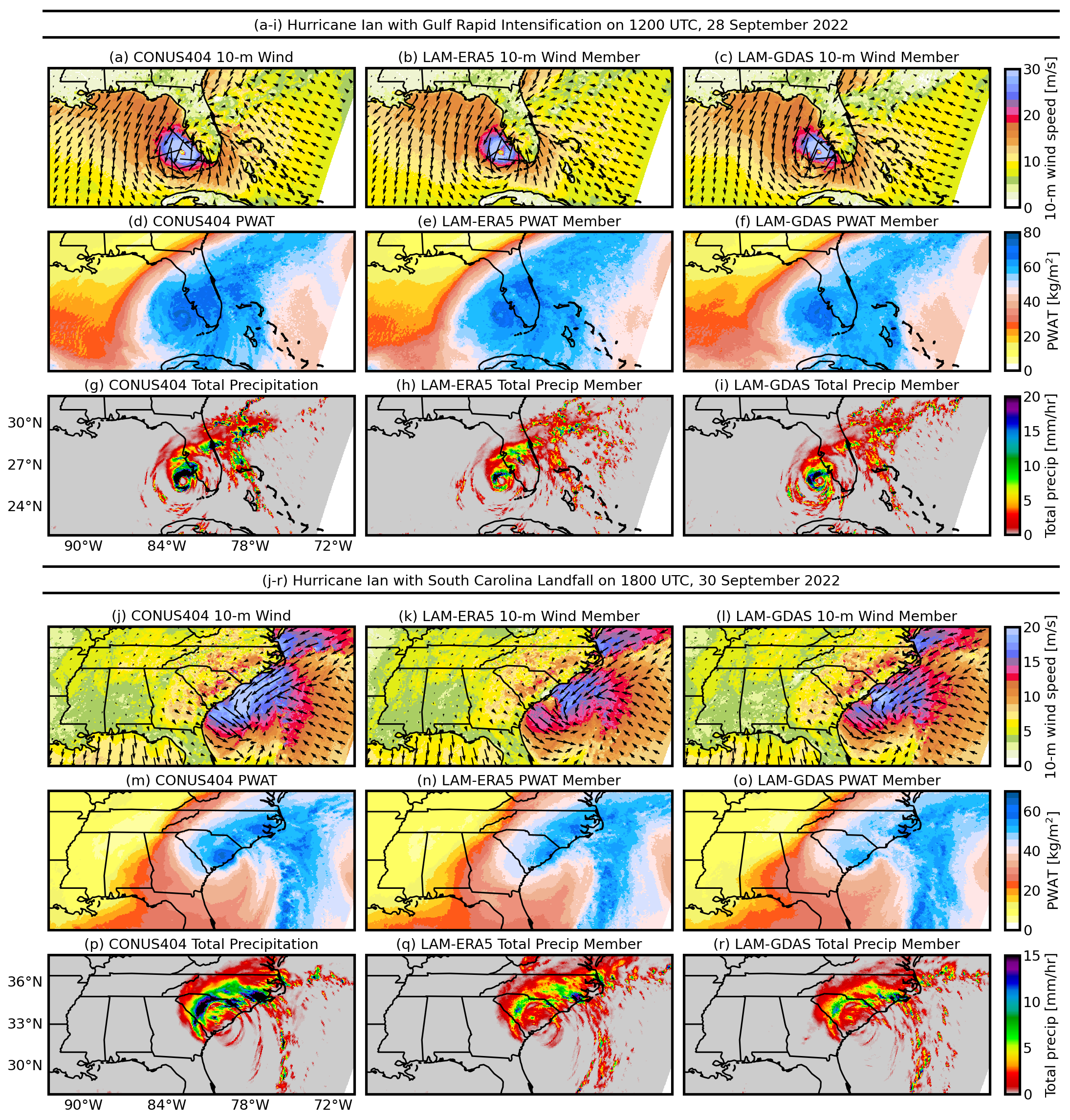}
    \caption{
    A case study of Hurricane Ian. 
    (a--c) 10-m wind arrows and wind speed as color shade for CONUS404, a LAM-ERA5 member, and a LAM-GDAS member, respectively, during the Gulf rapid intensification stage at 1200 UTC, 28 September 2022. 
    (d--f) As in (a--c), but for total precipitable water (``PWAT''). 
    (g--i) As in (a--c), but for total precipitation.
    (j--r) As in (a--i), but for the South Carolina landfall of Hurricane Ian at 1800 UTC, 30 September 2022.
    }
    \label{fig4}
\end{figure}

Case studies of Hurricane Ian are provided to further investigate the performance of the AI-based dynamical downscaling system on strong TCs (Figure~\ref{fig4}). At 1200~UTC 28~September 2022, Ian intensified rapidly over the eastern Gulf to its peak and hit southwestern Florida later that day, causing massive damage \cite{bucci2023ian}. In this case, the CONUS404 shows a closed eye with compact eyewall wind maxima, and gale-force flow toward the coast (Figure~\ref{fig4}a). LAM-ERA5 and LAM-GDAS members reproduce the eye diameter, eyewall placement, and the outer wind patterns correctly, with only small center-position offsets (Figures~\ref{fig4}b and \ref{fig4}c). For total precipitable water, LAM-ERA5 and LAM-GDAS members correctly capture the moist envelope around the core, the northeastward plume, and the sharp western gradient of wrapping dry air, pointing to a good downscaling performance of moisture distribution and transport (Figures~\ref{fig4}d--f). 

For total precipitation, the CONUS404 target shows an eyewall ring above 15 $\mathrm{mm\cdot hour^{-1}}$, spiral rainbands wrapping into the core, and scattered convective cells in the outer environment (Figure~\ref{fig4}g). LAM-ERA5 and LAM-GDAS members reproduce the ring pattern, rainband spacing and curvature, as well as the outer convective precipitation (Figures~\ref{fig4}h and i). Note that the individual cells are not collocated with the CONUS404, because the generative stage draws stochastic realizations, so agreement at these scales is not gridpoint-based. This result is consistent with the Figure~\ref{fig2} spectral verification.

By 1800~UTC 30~September, Ian made its final landfall on South Carolina, while undergoing extratropical transition \cite{bucci2023ian}. The CONUS404 pictures Ian with weaker and strongly asymmetric flows. The wind maxima are offshore, east of the center; onshore flow is directed at the coast, with a sharp contrast along the coastline. LAM-ERA5 and LAM-GDAS members reproduce these wind features well (Figures~\ref{fig4}j--l), together with the comma-shaped moisture pattern and the dry tail (Figures~\ref{fig4}m--o). For total precipitation, the CONUS404 target shows a broad inland rainband, with strong precipitation over west-central South Carolina and weaker offshore convective cells. These patterns are also well represented by LAM-ERA5 and LAM-GDAS members; their peak precipitation is only slightly weaker than the CONUS404 maximum (Figures~\ref{fig4}p--r).

Overall, the two case studies show that the system is capable of downscaling a strong TC for both its intensification over open water and weakening after landing. The downscaling performance is maintained across multiple LR boundary forcings and without case-specific tuning. This highlights the effectiveness and value of the AI-based dynamical downscaling system on TC applications.

\section{Discussion and conclusions}\label{sec4}

This study develops and evaluates a two-stage AI-based dynamical downscaling system for Atlantic TCs. An AI-based LAM downscales 3-hourly LR boundary forcings into hourly 8-km fields autoregressively, and a diffusion model converts the deterministic outputs into 20-member ensembles of hazard-relevant variables. Over the 2020--2024 Atlantic hurricane seasons, the system integrates stably without error spikes, and its performance largely improves over the deterministic Sha et al.~(2020) baseline. 

HURDAT2-based verifications show that the system inherits strengths from both its forcings and the CONUS404 training target. Its downscaled TC positions follow the large-scale forcings well, with several tracks closer to the HURDAT2 observations than CONUS404. The energy spectra analysis, on the other hand, confirms that the system can produce CONUS404-like spatial variance on storm-resolving scales. For TC extremes, the system is also skillful, with the largest improvements over the Sha et al.~(2020) baseline for extreme precipitation. Two case studies of Hurricane Ian illustrate these strengths, with TC eyewall winds, moisture patterns, and rainbands well produced during both the rapid intensification and decaying stages.

The system generalizes well to unseen forcings. From ERA5 to GDAS/FNL, skills for total precipitation and precipitable water are nearly unchanged; small gaps appear for 10-m wind and 2-m air temperature, most visibly for 15--30 $\mathrm{m\cdot s^{-1}}$ wind extremes, many of which can be explained by the ERA5-to-GDAS/FNL quality differences \cite{li2024exploring}. Since GDAS/FNL is produced by the GFS assimilation component, the system shows potential for applications driven by other GFS products, with possible fine-tuning to improve performance further.

A limitation of the system is that its ensemble spread comes solely from generative sampling, which yields miscalibrated extreme event probabilities; similar issues were found in other generative-AI-based works \cite<e.g. >{mardani2025residual,antonio2025postprocessing,guan2026high}. Future studies may consider LR forcing ensembles, diffusion model fine-tuning \cite{clark2024directly,jacq2026diffusion}, and the post-processing of diffusion model outputs \cite{asch2026rigorous,wang2026evaluating} as solutions. In addition, the system is trained over a fixed domain. Future work can fill this gap by training the system across multiple domains and incorporating large-scale flow-pattern-based guidance to improve spatial generalization.

In summary, this study proposes an AI-based dynamical downscaling system that effectively generates ensembles of hazard-relevant variables, supporting the risk assessment of Atlantic TCs and associated weather extremes. Taking this two-stage design as a basis, similar workflows can be extended to other hazard types and boundary forcing datasets, addressing broader challenges in estimating regional weather impacts.

%
%

\section*{Open Research Section}
The ERA5 reanalysis data for this study can be accessed through the NSF National Center for Atmospheric Research (NCAR) Geoscience Data Exchange (GDEX) \cite{ecmwf2019era5} and the Google Research, Analysis-Ready, Cloud Optimized (ARCO) ERA5 \cite{carver2023arcoera5}. CONUS404 data for water years 1980--2022 are available from the NSF NCAR Digital Assets Services Hub (DASH) \cite{CONUS404}. Readers may contact Dr. Lulin Xue and Dr. Aubrey Dugger via [conus404@ucar.edu] for the extended CONUS404 data for water years 2023--2024. The 0.25$^\circ$ GDAS/FNL is also available from DASH \cite{NCEP2015FNL}. The HURDAT2 dataset is available at \citeA{NHC2026HURDAT2}. The AI model simulation and verification code is available at \url{https://github.com/yingkaisha/RAL-GWC-TC/}.

\section*{Acknowledgments}
This material is based upon work supported by the National Science Foundation (NSF) National Center for Atmospheric Research (NCAR), which is a major facility sponsored by the U.S. National Science Foundation under Cooperative Agreement No. 1852977. 
Y. Sha and T. Mayo are also supported by the EdEC Faculty Innovator Program No. 1755088.
The authors acknowledge high-performance computing support from Derecho and Casper \cite{Cheyenne} provided by the Computational and Information Systems Laboratory, NCAR, and sponsored by the NSF.

\bibliographystyle{apacite} 
\bibliography{reference}

@article{bi2023accurate,
  title={Accurate medium-range global weather forecasting with 3D neural networks},
  author={Bi, Kaifeng and Xie, Lingxi and Zhang, Hengheng and Chen, Xin and Gu, Xiaotao and Tian, Qi},
  journal={Nature},
  volume={619},
  number={7970},
  pages={533--538},
  year={2023},
  publisher={Nature Publishing Group UK London}
}

@article{lam2023learning,
  title={Learning skillful medium-range global weather forecasting},
  author={Lam, Remi and Sanchez-Gonzalez, Alvaro and Willson, Matthew and Wirnsberger, Peter and Fortunato, Meire and Alet, Ferran and Ravuri, Suman and Ewalds, Timo and Eaton-Rosen, Zach and Hu, Weihua and others},
  journal={Science},
  volume={382},
  number={6677},
  pages={1416--1421},
  year={2023},
  publisher={American Association for the Advancement of Science}
}

@article{chen2023fuxi,
  title={FuXi: A cascade machine learning forecasting system for 15-day global weather forecast},
  author={Chen, Lei and Zhong, Xiaohui and Zhang, Feng and Cheng, Yuan and Xu, Yinghui and Qi, Yuan and Li, Hao},
  journal={npj Climate and Atmospheric Science},
  volume={6},
  number={1},
  pages={190},
  year={2023},
  publisher={Nature Publishing Group UK London}
}

@article{hersbach2020era5,
  title={The ERA5 global reanalysis},
  author={Hersbach, Hans and Bell, Bill and Berrisford, Paul and Hirahara, Shoji and Hor{\'a}nyi, Andr{\'a}s and Mu{\~n}oz-Sabater, Joaqu{\'\i}n and Nicolas, Julien and Peubey, Carole and Radu, Raluca and Schepers, Dinand and others},
  journal={Quarterly Journal of the Royal Meteorological Society},
  volume={146},
  number={730},
  pages={1999--2049},
  year={2020},
  publisher={Wiley Online Library}
}

@article{watt2023ace,
  title={{ACE}: A fast, skillful learned global atmospheric model for climate prediction},
  author={Watt-Meyer, Oliver and Dresdner, Gideon and McGibbon, Jeremy and Clark, Spencer K and Henn, Brian and Duncan, James and Brenowitz, Noah D and Kashinath, Karthik and Pritchard, Michael S and Bonev, Boris and others},
  journal={arXiv preprint arXiv:2310.02074},
  year={2023}
}

@misc{ecmwf2019era5,
  author       = {{European Centre for Medium-Range Weather Forecasts}},
  title        = {{ERA5 Reanalysis (0.25 Degree Latitude-Longitude Grid) [Dataset]}},
  year         = {2019},
  publisher    = {Geoscience Data Exchange at the National Center for Atmospheric Research, Computational and Information Systems Laboratory},
  doi          = {10.5065/BH6N-5N20},
  url          = {https://doi.org/10.5065/BH6N-5N20},
  note         = {Accessed 09 September 2025}
}

@misc{CONUS404,
  author       = {Rasmussen, Roy M. and Liu, Changhai and Ikeda, Kyoko and Chen, Fei and Kim, Jongsang and Schneider, Tiffany and Gochis, David and Dugger, Andrew and Viger, Roland},
  title        = {{Four-kilometer long-term regional hydroclimate reanalysis over the conterminous United States (CONUS) [Dataset]}},
  year         = {2023},
  publisher    = {{NSF National Center for Atmospheric Research}},
  doi          = {10.5065/ZYY0-Y036},
  url          = {https://doi.org/10.5065/ZYY0-Y036},
  note         = {Accessed 09 September 2025}
}

@misc{NCEP2015FNL,
  author       = {{National Centers for Environmental Prediction}},
  title        = {{NCEP GDAS/FNL 0.25 Degree Global Tropospheric Analyses and Forecast Grids [Dataset]}},
  year         = {2015},
  publisher    = {{National Weather Service, NOAA, U.S. Department of Commerce; NSF National Center for Atmospheric Research}},
  note         = {Accessed 01 March 2026},
  doi          = {10.5065/D65Q4T4Z},
  url          = {https://doi.org/10.5065/D65Q4T4Z}
}

@misc{NHC2026HURDAT2,
  author       = {{National Hurricane Center}},
  title        = {{Atlantic Hurricane Database (HURDAT2) [Dataset]}},
  year         = {2026},
  publisher    = {{National Weather Service, NOAA, U.S. Department of Commerce}},
  note         = {Accessed 01 March 2026},
  url          = {https://www.nhc.noaa.gov/data/}
}

@inproceedings{carver2023arcoera5,
  author = {Carver, Robert W and Merose, Alex},
  title = {{ARCO-ERA5: An Analysis-Ready Cloud-Optimized Reanalysis Dataset [Dataset]}},
  booktitle = {22nd Conference on AI for Environmental Science},
  year = {2023},
  note = {Accessed on 01 Sep 2024},
  location = {Denver, CO},
  publisher = {American Meteorological Society},
  paper = {4A.1},
  url = {https://ams.confex.com/ams/103ANNUAL/meetingapp.cgi/Paper/415842}
}

@techreport{Cheyenne,
	author = {{Computational and Information Systems Laboratory, CISL}},
	institution = {{National Center for Atmospheric Research}},
	title = {{Cheyenne: HPE/SGI ICE XA System (NCAR Community Computing)}},
	url = {https://doi.org/10.5065/D6RX99HX},
	year = {2020}}

@inproceedings{paszke2019pytorch,
  title={PyTorch: An Imperative Style, High-Performance Deep Learning Library},
  author={Paszke, Adam and Gross, Sam and Massa, Francisco and Lerer, Adam and Bradbury, James and Chanan, Gregory and Killeen, Trevor and Lin, Zeming and Gimelshein, Natalia and Antiga, Luca and Desmaison, Alban and K{\"o}pf, Andreas and Yang, Edward and DeVito, Zachary and Raison, Martin and Tejani, Alykhan and Chilamkurthy, Sasank and Steiner, Benoit and Fang, Lu and Bai, Junjie and Chintala, Soumith},
  booktitle={Advances in Neural Information Processing Systems},
  pages={8024--8035},
  year={2019}
}

@article{bodnar2024aurora,
  title={Aurora: A foundation model of the atmosphere},
  author={Bodnar, Cristian and Bruinsma, Wessel P and Lucic, Ana and Stanley, Megan and Brandstetter, Johannes and Garvan, Patrick and Riechert, Maik and Weyn, Jonathan and Dong, Haiyu and Vaughan, Anna and others},
  journal={arXiv preprint arXiv:2405.13063},
  year={2024}
}

@article{lang2024aifs,
  title={{AIFS-ECMWF}'s data-driven forecasting system},
  author={Lang, Simon and Alexe, Mihai and Chantry, Matthew and Dramsch, Jesper and Pinault, Florian and Raoult, Baudouin and Clare, Mariana CA and Lessig, Christian and Maier-Gerber, Michael and Magnusson, Linus and others},
  journal={arXiv preprint arXiv:2406.01465},
  year={2024}
}

@article{sha2025improving,
  title={Improving AI weather prediction models using global mass and energy conservation schemes},
  author={Sha, Yingkai and Schreck, John S and Chapman, William and Gagne, David John},
  journal={Journal of Advances in Modeling Earth Systems},
  volume={17},
  number={11},
  pages={e2025MS005138},
  year={2025},
  publisher={Wiley Online Library}
}

@article{price2025probabilistic,
  title={Probabilistic weather forecasting with machine learning},
  author={Price, Ilan and Sanchez-Gonzalez, Alvaro and Alet, Ferran and Andersson, Tom R and El-Kadi, Andrew and Masters, Dominic and Ewalds, Timo and Stott, Jacklynn and Mohamed, Shakir and Battaglia, Peter and others},
  journal={Nature},
  volume={637},
  number={8044},
  pages={84--90},
  year={2025},
  publisher={Nature Publishing Group}
}

@article{nipen2025regional,
  title={Regional data-driven weather modeling with a global stretched-grid},
  author={Nipen, Thomas Nils and Haugen, H{\aa}vard Homleid and Ingstad, Magnus Sikora and Nordhagen, Even Marius and Salihi, Aram Farhad Shafiq and Tedesco, Paulina and Seierstad, Ivar Ambj{\o}rn and Kristiansen, J{\o}rn and Lang, Simon and Alexe, Mihai and others},
  journal={Artificial Intelligence for the Earth Systems},
  year={2025},
  publisher={American Meteorological Society}
}

@article{abdi2025hrrrcast,
  title={{HRRRCast}: a data-driven emulator for regional weather forecasting at convection allowing scales},
  author={Abdi, Daniel and Jankov, Isidora and Madden, Paul and Vargas, Vanderlei and Smith, Timothy A and Frolov, Sergey and Flora, Montgomery and Potvin, Corey},
  journal={arXiv preprint arXiv:2507.05658},
  year={2025}
}

@article{xu2025artificial,
  title={An artificial intelligence-based limited area model for forecasting of surface meteorological variables},
  author={Xu, Pengbo and Zheng, Xiaogu and Gao, Tianyan and Wang, Yu and Yin, Junping and Zhang, Juan and Zhang, Xuanze and Luo, San and Wang, Zhonglei and Zhang, Zhimin and others},
  journal={Communications Earth \& Environment},
  volume={6},
  number={1},
  pages={372},
  year={2025},
  publisher={Nature Publishing Group UK London}
}

@article{prein2015review,
  title={A review on regional convection-permitting climate modeling: Demonstrations, prospects, and challenges},
  author={Prein, Andreas F and Langhans, Wolfgang and Fosser, Giorgia and Ferrone, Andrew and Ban, Nikolina and Goergen, Klaus and Keller, Michael and T{\"o}lle, Merja and Gutjahr, Oliver and Feser, Frauke and others},
  journal={Reviews of geophysics},
  volume={53},
  number={2},
  pages={323--361},
  year={2015},
  publisher={Wiley Online Library}
}

@article{rasmussen2023conus404,
  title={CONUS404: The NCAR--USGS 4-km long-term regional hydroclimate reanalysis over the CONUS},
  author={Rasmussen, RM and Chen, Fei and Liu, CH and Ikeda, Kyoko and Prein, A and Kim, J and Schneider, T and Dai, A and Gochis, D and Dugger, A and others},
  journal={Bulletin of the American Meteorological Society},
  volume={104},
  number={8},
  pages={E1382--E1408},
  year={2023},
  publisher={American Meteorological Society}
}

@article{loshchilov2017decoupled,
  title={Decoupled weight decay regularization},
  author={Loshchilov, Ilya and Hutter, Frank},
  journal={arXiv preprint arXiv:1711.05101},
  year={2017}
}

@article{schreck2025community,
  title={Community Research Earth Digital Intelligence Twin: a scalable framework for AI-driven Earth System Modeling},
  author={Schreck, John S and Sha, Yingkai and Chapman, William and Kimpara, Dhamma and Berner, Judith and McGinnis, Seth and Kazadi, Arnold and Sobhani, Negin and Kirk, Ben and Becker, Charlie and others},
  journal={npj Climate and Atmospheric Science},
  volume={8},
  number={1},
  pages={239},
  year={2025},
  publisher={Nature Publishing Group UK London}
}

@article{sha2025investigating,
  title={Investigating the Use of Terrain-Following Coordinates in AI-Driven Precipitation Forecasts},
  author={Sha, Yingkai and Schreck, John S and Chapman, William and Gagne, David John},
  journal={Geophysical Research Letters},
  volume={52},
  number={20},
  pages={e2025GL118478},
  year={2025},
  publisher={Wiley Online Library}
}

@article{morss2009spectra,
  title={Spectra, spatial scales, and predictability in a quasigeostrophic model},
  author={Morss, Rebecca E and Snyder, Chris and Rotunno, Richard},
  journal={Journal of the atmospheric sciences},
  volume={66},
  number={10},
  pages={3115--3130},
  year={2009}
}

@article{schaefer1990critical,
  title={The critical success index as an indicator of warning skill},
  author={Schaefer, Joseph T},
  journal={Weather and forecasting},
  volume={5},
  number={4},
  pages={570--575},
  year={1990}
}

@article{wang2024set,
  title={How to set {AdamW}'s weight decay as you scale model and dataset size},
  author={Wang, Xi and Aitchison, Laurence},
  journal={arXiv preprint arXiv:2405.13698},
  year={2024}
}

@article{sha2020adeep,
  title={Deep-learning-based gridded downscaling of surface meteorological variables in complex terrain. Part {I}: Daily maximum and minimum 2-m temperature},
  author={Sha, Yingkai and Gagne II, David John and West, Gregory and Stull, Roland},
  journal={Journal of Applied Meteorology and Climatology},
  volume={59},
  number={12},
  pages={2057--2073},
  year={2020}
}

@article{sha2020bdeep,
  title={Deep-learning-based gridded downscaling of surface meteorological variables in complex terrain. Part {II}: Daily precipitation},
  author={Sha, Yingkai and Gagne II, David John and West, Gregory and Stull, Roland},
  journal={Journal of Applied Meteorology and Climatology},
  volume={59},
  number={12},
  pages={2075--2092},
  year={2020}
}

@article{mardani2025residual,
  title   = {Residual corrective diffusion modeling for km-scale atmospheric downscaling},
  author  = {Mardani, Morteza and Brenowitz, Noah and Cohen, Yair and Pathak, Jaideep and Chen, Chieh-Yu and Liu, Cheng-Chin and Vahdat, Arash and Nabian, Mohammad Amin and Ge, Tao and Subramaniam, Akshay and Kashinath, Karthik and Kautz, Jan and Pritchard, Mike},
  journal = {Communications Earth \& Environment},
  volume  = {6},
  pages   = {124},
  year    = {2025},
  doi     = {10.1038/s43247-025-02042-5}
}

@article{pathak2026kilometer,
  title   = {Kilometer-scale convection-allowing model emulation using generative diffusion modeling},
  author  = {Pathak, Jaideep and Cohen, Yair and Garg, Piyush and Harrington, Peter and Brenowitz, Noah and Durran, Dale and Mardani, Morteza and Vahdat, Arash and Xu, Shaoming and Kashinath, Karthik and Pritchard, Michael},
  journal = {Science Advances},
  volume  = {12},
  number  = {5},
  pages   = {eadv0423},
  year    = {2026},
  doi     = {10.1126/sciadv.adv0423}
}

@article{vosper2023deep,
  title   = {Deep Learning for Downscaling Tropical Cyclone Rainfall to Hazard-Relevant Spatial Scales},
  author  = {Vosper, Emily and Watson, Peter and Harris, Lucy and McRae, Andrew and Santos-Rodriguez, Raul and Aitchison, Laurence and Mitchell, Dann},
  journal = {Journal of Geophysical Research: Atmospheres},
  volume  = {128},
  number  = {10},
  pages   = {e2022JD038163},
  year    = {2023},
  doi     = {10.1029/2022JD038163}
}

@article{emanuel2005increasing,
  title={Increasing destructiveness of tropical cyclones over the past 30 years},
  author={Emanuel, Kerry},
  journal={Nature},
  volume={436},
  number={7051},
  pages={686--688},
  year={2005},
  publisher={Nature Publishing Group}
}

@article{knutson2020tropical,
  title   = {Tropical Cyclones and Climate Change Assessment: {P}art {II}: {P}rojected Response to Anthropogenic Warming},
  author={Knutson, Thomas and Camargo, Suzana J and Chan, Johnny CL and Emanuel, Kerry and Ho, Chang-Hoi and Kossin, James and Mohapatra, Mrutyunjay and Satoh, Masaki and Sugi, Masato and Walsh, Kevin and others},
  journal={Bulletin of the American Meteorological Society},
  volume={101},
  number={3},
  pages={E303--E322},
  year={2020},
  publisher={American Meteorological Society}
}

@article{roberts2020impact,
  title={Impact of model resolution on tropical cyclone simulation using the {HighResMIP--PRIMAVERA} multimodel ensemble},
  author={Roberts, Malcolm John and Camp, Joanne and Seddon, Jon and Vidale, Pier Luigi and Hodges, Kevin and Vanniere, Benoit and Mecking, Jenny and Haarsma, Rein and Bellucci, Alessio and Scoccimarro, Enrico and others},
  journal={Journal of Climate},
  volume={33},
  number={7},
  pages={2557--2583},
  year={2020}
}

@article{gutmann2018changes,
  title={Changes in hurricanes from a 13-yr convection-permitting pseudo--global warming simulation},
  author={Gutmann, E. and Rasmussen, Roy M and Liu, Changhai and Ikeda, Kyoko and Bruyere, Cindy L and Done, James M and Garr{\`e}, Luca and Friis-Hansen, Peter and Veldore, Vidyunmala},
  journal={Journal of Climate},
  volume={31},
  number={9},
  pages={3643--3657},
  year={2018}
}

@article{emanuel2006statistical,
  title={A statistical deterministic approach to hurricane risk assessment},
  author={Emanuel, Kerry and Ravela, Sai and Vivant, Emmanuel and Risi, Camille},
  journal={Bulletin of the American Meteorological Society},
  volume={87},
  number={3},
  pages={299--314},
  year={2006},
  publisher={American Meteorological Society}
}

@article{karras2022elucidating,
  title={Elucidating the design space of diffusion-based generative models},
  author={Karras, Tero and Aittala, Miika and Aila, Timo and Laine, Samuli},
  journal={Advances in neural information processing systems},
  volume={35},
  pages={26565--26577},
  year={2022}
}

@inproceedings{perez2018film,
  title={{Film}: Visual reasoning with a general conditioning layer},
  author={Perez, Ethan and Strub, Florian and De Vries, Harm and Dumoulin, Vincent and Courville, Aaron},
  booktitle={Proceedings of the AAAI conference on artificial intelligence},
  volume={32},
  number={1},
  year={2018}
}

@inproceedings{vandaldeepsd2017,
    address = {Halifax, NS, Canada},
    title = {{DeepSD}: generating high resolution climate change projections through single image super-resolution},
    isbn = {978-1-4503-4887-4},
    shorttitle = {{DeepSD}},
    url = {http://dl.acm.org/citation.cfm?doid=3097983.3098004},
    doi = {10.1145/3097983.3098004},
    language = {en},
    urldate = {2019-12-30},
    booktitle = {Proceedings of the 23rd {ACM} {SIGKDD} {International} {Conference} on {Knowledge} {Discovery} and {Data} {Mining} - {KDD} '17},
    publisher = {ACM Press},
    author = {Vandal, Thomas and Kodra, Evan and Ganguly, Sangram and Michaelis, Andrew and Nemani, Ramakrishna and Ganguly, Auroop R.},
    year = {2017},
    pages = {1663--1672},
}

@techreport{bucci2023ian,
  title       = {Tropical Cyclone Report: Hurricane Ian (AL092022)},
  author      = {Bucci, Lisa and Alaka, Laura and Hagen, Andrew and Delgado, Sandy and Beven, John},
  institution = {NOAA/NWS National Hurricane Center},
  year        = {2023},
  url         = {https://www.nhc.noaa.gov/data/tcr/AL092022_Ian.pdf}
}

@article{maraun2010precipitation,
  title   = {Precipitation downscaling under climate change: {R}ecent developments to bridge the gap between dynamical models and the end user},
  author  = {Maraun, Douglas and Wetterhall, Fredrik and Ireson, Andrew M. and Chandler, Richard E. and Kendon, Elizabeth J. and Widmann, Martin and Brienen, Susanne and Rust, Henning W. and Sauter, Thomas and Themessl, Matthias and Venema, Victor K. C. and Chun, Kwok P. and Goodess, Clare M. and Jones, Richard G. and Onof, Christian and Vrac, Mathieu and Thiele-Eich, Insa},
  journal = {Reviews of Geophysics},
  volume  = {48},
  number  = {3},
  pages   = {RG3003},
  year    = {2010},
  doi     = {10.1029/2009RG000314}
}

@article{gutmann2014intercomparison,
  title   = {An intercomparison of statistical downscaling methods used for water resource assessments in the {U}nited {S}tates},
  author  = {Gutmann, E. and Pruitt, Tom and Clark, Martyn P. and Brekke, Levi and Arnold, Jeffrey R. and Raff, David A. and Rasmussen, Roy M.},
  journal = {Water Resources Research},
  volume  = {50},
  number  = {9},
  pages   = {7167--7186},
  year    = {2014},
  doi     = {10.1002/2014WR015559}
}

@article{harris2022generative,
  title   = {A Generative Deep Learning Approach to Stochastic Downscaling of Precipitation Forecasts},
  author  = {Harris, Lucy and McRae, Andrew T. T. and Chantry, Matthew and Dueben, Peter D. and Palmer, Tim N.},
  journal = {Journal of Advances in Modeling Earth Systems},
  volume  = {14},
  number  = {10},
  pages   = {e2022MS003120},
  year    = {2022},
  doi     = {10.1029/2022MS003120}
}

@article{sha2026regional,
  title   = {AI-Based Regional Emulation for Kilometer-Scale Dynamical Downscaling},
  author  = {Sha, Yingkai and Hertneky, Tracy and Gutmann, Ethan D. and McGinnis, Seth and McCrary, Rachel and Xue, Lulin and Gagne II, David John and Newman, Kathryn and Newman, Andrew},
  journal = {arXiv preprint arXiv:2602.18646},
  year    = {2026}
}

@article{klotzbach2018continental,
  title   = {Continental {U.S.} Hurricane Landfall Frequency and Associated Damage: Observations and Future Risks},
  author  = {Klotzbach, Philip J. and Bowen, Steven G. and Pielke, Roger, Jr. and Bell, Michael},
  journal = {Bulletin of the American Meteorological Society},
  volume  = {99},
  number  = {7},
  pages   = {1359--1376},
  year    = {2018},
  doi     = {10.1175/BAMS-D-17-0184.1}
}

@article{landsea2013atlantic,
  title   = {Atlantic Hurricane Database Uncertainty and Presentation of a New Database Format},
  author  = {Landsea, Christopher W. and Franklin, James L.},
  journal = {Monthly Weather Review},
  volume  = {141},
  number  = {10},
  pages   = {3576--3592},
  year    = {2013},
  doi     = {10.1175/MWR-D-12-00254.1}
}

@article{hersbach2000decomposition,
  title   = {Decomposition of the Continuous Ranked Probability Score for Ensemble Prediction Systems},
  author  = {Hersbach, Hans},
  journal = {Weather and Forecasting},
  volume  = {15},
  number  = {5},
  pages   = {559--570},
  year    = {2000},
  doi     = {10.1175/1520-0434(2000)015<0559:DOTCRP>2.0.CO;2}
}

@article{gneiting2007strictly,
  title   = {Strictly Proper Scoring Rules, Prediction, and Estimation},
  author  = {Gneiting, Tilmann and Raftery, Adrian E.},
  journal = {Journal of the American Statistical Association},
  volume  = {102},
  number  = {477},
  pages   = {359--378},
  year    = {2007},
  doi     = {10.1198/016214506000001437}
}

@article{skamarock2004evaluating,
  title   = {Evaluating Mesoscale {NWP} Models Using Kinetic Energy Spectra},
  author  = {Skamarock, William C.},
  journal = {Monthly Weather Review},
  volume  = {132},
  number  = {12},
  pages   = {3019--3032},
  year    = {2004},
  doi     = {10.1175/MWR2830.1}
}

@article{ferro2011extremal,
  title   = {Extremal Dependence Indices: Improved Verification Measures for Deterministic Forecasts of Rare Binary Events},
  author  = {Ferro, Christopher A. T. and Stephenson, David B.},
  journal = {Weather and Forecasting},
  volume  = {26},
  number  = {5},
  pages   = {699--713},
  year    = {2011},
  doi     = {10.1175/WAF-D-10-05030.1}
}

@article{roberts2008scale,
  title   = {Scale-Selective Verification of Rainfall Accumulations from High-Resolution Forecasts of Convective Events},
  author  = {Roberts, Nigel M. and Lean, Humphrey W.},
  journal = {Monthly Weather Review},
  volume  = {136},
  number  = {1},
  pages   = {78--97},
  year    = {2008},
  doi     = {10.1175/2007MWR2123.1}
}

@book{wilks2019statistical,
  title     = {Statistical Methods in the Atmospheric Sciences},
  author    = {Wilks, Daniel S.},
  edition   = {4th},
  publisher = {Elsevier},
  year      = {2019}
}

@inproceedings{ho2020denoising,
  author    = {Ho, Jonathan and Jain, Ajay and Abbeel, Pieter},
  title     = {Denoising diffusion probabilistic models},
  booktitle = {Advances in Neural Information Processing Systems},
  volume    = {33},
  pages     = {6840--6851},
  year      = {2020}
}

@article{camargo2016tropical,
  title={Tropical cyclones in climate models},
  author={Camargo, Suzana J and Wing, Allison A},
  journal={Wiley Interdisciplinary Reviews: Climate Change},
  volume={7},
  number={2},
  pages={211--237},
  year={2016},
  publisher={Wiley Online Library}
}

@article{lucas2021convection,
  title={Convection-permitting modeling with regional climate models: Latest developments and next steps},
  author={Lucas-Picher, Philippe and Arg{\"u}eso, Daniel and Brisson, Erwan and Tramblay, Yves and Berg, Peter and Lemonsu, Aude and Kotlarski, Sven and Caillaud, C{\'e}cile},
  journal={Wiley Interdisciplinary Reviews: Climate Change},
  volume={12},
  number={6},
  pages={e731},
  year={2021},
  publisher={Wiley Online Library}
}

@article{bano2020configuration,
  title={Configuration and intercomparison of deep learning neural models for statistical downscaling},
  author={Ba{\~n}o-Medina, Jorge and Manzanas, Rodrigo and Guti{\'e}rrez, Jos{\'e} Manuel},
  journal={Geoscientific Model Development},
  volume={13},
  number={4},
  pages={2109--2124},
  year={2020},
  publisher={Copernicus Publications G{\"o}ttingen, Germany}
}

@article{miralles2022downscaling,
  title={Downscaling of historical wind fields over {Switzerland} using generative adversarial networks},
  author={Miralles, Oph{\'e}lia and Steinfeld, Daniel and Martius, Olivia and Davison, Anthony C},
  journal={Artificial Intelligence for the Earth Systems},
  volume={1},
  number={4},
  pages={e220018},
  year={2022}
}

@article{antonio2025postprocessing,
  title={Postprocessing East African rainfall forecasts using a generative machine learning model},
  author={Antonio, Bobby and McRae, Andrew TT and MacLeod, David and Cooper, Fenwick C and Marsham, John and Aitchison, Laurence and Palmer, Tim N and Watson, Peter AG},
  journal={Journal of Advances in Modeling Earth Systems},
  volume={17},
  number={3},
  pages={e2024MS004796},
  year={2025},
  publisher={Wiley Online Library}
}

@article{guan2026high,
  title={High-Resolution Climate Projections Using Diffusion-Based Downscaling of a Lightweight Climate Emulator},
  author={Guan, Haiwen and Chakraborty, Dibyajyoti and Darman, Moein and Arcomano, Troy and Chattopadhyay, Ashesh and Maulik, Romit},
  journal={arXiv preprint arXiv:2602.13416},
  year={2026}
}

@inproceedings{shi2016real,
  title={Real-time single image and video super-resolution using an efficient sub-pixel convolutional neural network},
  author={Shi, Wenzhe and Caballero, Jose and Husz{\'a}r, Ferenc and Totz, Johannes and Aitken, Andrew P and Bishop, Rob and Rueckert, Daniel and Wang, Zehan},
  booktitle={Proceedings of the IEEE conference on computer vision and pattern recognition},
  pages={1874--1883},
  year={2016}
}

@article{li2024exploring,
  title={Exploring the differences in kinetic energy spectra between the {NCEP FNL} and {ERA5} datasets},
  author={Li, Zongheng and Peng, Jun and Zhang, Lifeng and Guan, Jiping},
  journal={Journal of the Atmospheric Sciences},
  volume={81},
  number={2},
  pages={363--380},
  year={2024},
  publisher={American Meteorological Society}
}

@article{jacq2026diffusion,
  title={Diffusion Fine-tuning with Rewarded Moment Matching Distillation},
  author={Jacq, Alexis and Couairon, Guillaume and De Bortoli, Valentin and Berthet, Quentin and Doucet, Arnaud and Elie, Romuald},
  journal={arXiv preprint arXiv:2606.30414},
  year={2026}
}

@inproceedings{clark2024directly,
  title={Directly fine-tuning diffusion models on differentiable rewards},
  author={Clark, Kevin and Vicol, Paul and Swersky, Kevin and Fleet, David},
  booktitle={International Conference on Learning Representations},
  volume={2024},
  pages={4793--4822},
  year={2024}
}

@article{asch2026rigorous,
  title={Rigorous uncertainty quantification of probabilistic AI weather forecasts with conformal prediction},
  author={Asch, Anna and Rossellini, Raphael and Hassanzadeh, Pedram and Willett, Rebecca},
  journal={arXiv preprint arXiv:2606.19642},
  year={2026}
}

@article{wang2026evaluating,
  title={Evaluating and Calibrating Diffusion Model-derived Uncertainty for Quantitative MRI Mapping},
  author={Wang, Shishuai and Klein, Stefan and Hernandez-Tamames, Juan A and Poot, Dirk HJ},
  journal={arXiv preprint arXiv:2608.11942},
  year={2026}
}

@article{sun2024deep,
  title={Deep learning in statistical downscaling for deriving high spatial resolution gridded meteorological data: A systematic review},
  author={Sun, Yongjian and Deng, Kefeng and Ren, Kaijun and Liu, Jia and Deng, Chongjiu and Jin, Yongjun},
  journal={ISPRS Journal of Photogrammetry and Remote Sensing},
  volume={208},
  pages={14--38},
  year={2024},
  publisher={Elsevier}
}

@article{lopez2025dynamical,
  title={Dynamical-generative downscaling of climate model ensembles},
  author={Lopez-Gomez, Ignacio and Wan, Zhong Yi and Zepeda-N{\'u}{\~n}ez, Leonardo and Schneider, Tapio and Anderson, John and Sha, Fei},
  journal={Proceedings of the National Academy of Sciences},
  volume={122},
  number={17},
  pages={e2420288122},
  year={2025},
  publisher={National Academy of Sciences}
}

@article{wang2021deep,
  title={Deep learning for daily precipitation and temperature downscaling},
  author={Wang, Fang and Tian, Di and Lowe, Lisa and Kalin, Latif and Lehrter, John},
  journal={Water Resources Research},
  volume={57},
  number={4},
  pages={e2020WR029308},
  year={2021},
  publisher={Wiley Online Library}
}

@incollection{hsieh2023deep,
  author    = {Hsieh, William W.},
  title     = {Deep Learning},
  booktitle = {Introduction to Environmental Data Science},
  publisher = {Cambridge University Press},
  address   = {Cambridge, UK},
  pages     = {494--517},
  year      = {2023},
  doi       = {10.1017/9781107588493.016}}

@article{powell2007tropical,
  title = {Tropical cyclone destructive potential by integrated kinetic energy},
  author = {Powell, Mark D. and Reinhold, Timothy A.},
  journal = {Bulletin of the American Meteorological Society},
  volume = {88},
  number = {4},
  pages = {513--526},
  year = {2007},
  doi = {10.1175/BAMS-88-4-513}
}

\end{document}